\documentclass[nofootinbib,twocolumn,showpacs,preprintnumbers,pre,aps]{revtex4-1}

\usepackage[dvipdfmx]{graphicx}
\usepackage{bm}
\usepackage{amsmath}
\usepackage{amssymb}
\usepackage{color}
\usepackage{here}
\usepackage{siunitx}

\begin{document}

\title{Non-equilibrium probability flux of a thermally driven micromachine}%

\author{Isamu Sou}

\author{Yuto Hosaka}

\author{Kento Yasuda}

\author{Shigeyuki Komura}
\email{komura@tmu.ac.jp}

\affiliation{
Department of Chemistry, Graduate School of Science,
Tokyo Metropolitan University, Tokyo 192-0397, Japan}

\date{\today}

\begin{abstract}
We discuss the non-equilibrium statistical mechanics of a thermally driven micromachine consisting 
of three spheres and two harmonic springs [Y. Hosaka \textit{et al.},  J.\ Phys.\ Soc.\ Jpn.\ 
\textbf{86}, 113801 (2017)]. 
We obtain the non-equilibrium steady state probability distribution function of such 
a micromachine and calculate its probability flux in the corresponding configuration space. 
The resulting probability flux can be expressed in terms of a frequency matrix that is used to 
distinguish between a non-equilibrium steady state and a thermal equilibrium state satisfying 
detailed balance.
The frequency matrix is shown to be proportional to the temperature difference between the spheres.
We obtain a linear relation between the eigenvalue of the frequency matrix and the average 
velocity of a thermally driven micromachine that can undergo a directed motion in a viscous fluid.
This relation is consistent with the scallop theorem for a deterministic three-sphere microswimmer.
\end{abstract}

\maketitle

\section{Introduction}
\label{sec:introduction}

Microswimmers are tiny machines that swim in a fluid and they are expected to be used in microfluidics 
and microsystems~\cite{Lauga09}.
Over the length scale of microswimmers, the fluid forces acting on them are dominated by the frictional 
viscous forces.
By transforming chemical energy into mechanical energy, however, microswimmers change their shape 
and move efficiently in viscous environments.
According to Purcell's scallop theorem, time-reversible body motion cannot be 
used for locomotion in a Newtonian fluid~\cite{Purcell77,Lauga11}.
As one of the simplest models exhibiting broken time-reversal symmetry, 
Najafi and Golestanian proposed a three-sphere swimmer~\cite{Golestanian04,Golestanian08}, 
in which three in-line spheres are linked by two arms of varying length.
Recently, such a swimmer has been experimentally realized by using colloidal beads manipulated by 
optical tweezers~\cite{Leoni09} 
or by controlling ferromagnetic particles at an air-water 
interface~\cite{Grosjean16,Grosjean18}.

\begin{figure}[tbh]
\begin{center}
\includegraphics[scale=0.35]{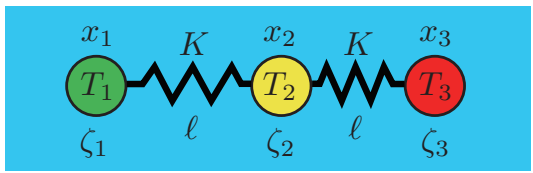}
\end{center}
\caption{(Color online)　
Thermally driven elastic three-sphere micromachine.
Three spheres are connected by two identical harmonic springs characterized by the elastic 
constant $K$ and the natural length $\ell$.
The time-dependent positions of the spheres are denoted by $x_i(t)$ ($i=1, 2, 3$) in a 
one-dimensional coordinate system, and $\zeta_i$ is the friction coefficient for $i$-th sphere.
The three spheres are in equilibrium with independent heat baths at temperatures $T_i$.
In this paper, we do not take into account hydrodynamic interactions acting between different 
spheres.}
\label{model}
\end{figure}

Recently, the present authors have proposed a generalized three-sphere microswimmer model 
in which the spheres are connected by two harmonic springs, i.e., an elastic 
microswimmer~\cite{Yasuda17,Kuroda19}.  
A similar model was also considered by other people~\cite{Dunkel09,Pande15,Pande17}.
Later, our model was further extended to a thermally driven elastic microswimmers~\cite{Hosaka17}, 
suggesting a new mechanism for locomotion that is purely induced by thermal fluctuations
without any external forcing.
As depicted in Fig.~\ref{model}, the key assumption is that the three spheres are 
in equilibrium with independent heat baths characterized by different temperatures.
In such a situation, heat transfer occurs inside the micromachine from a hotter sphere to a colder one, 
driving the whole system out of equilibrium. 
We have shown that a combination of heat transfer and hydrodynamic interactions 
among the spheres can lead to directional locomotion in a steady state~\cite{Hosaka17}. 
Our model has a similarity to a class of thermal ratchet 
models~\cite{Magnasco93,Astumian94,Rousselet94},
and the suggested new mechanism is relevant to non-equilibrium dynamics of proteins and enzymes 
in biological systems~\cite{Jee18,Dey19}.

For a thermally driven elastic micromachine, the average velocity was calculated to be~\cite{Hosaka17} 
\begin{align}
\langle V\rangle=\frac{k_{\rm B}(T_3-T_1)}{96\pi\eta\ell^2},
\label{Vsymmetric}
\end{align}
where $k_{\rm B}$ is the Boltzmann constant, $T_1$ and $T_3$ are the temperatures 
of the first and the third spheres (see Fig.~\ref{model}), $\eta$ is the viscosity of the 
surrounding fluid, and $\ell$ is the natural length of the two springs.
This result indicates that the swimming direction is from a colder sphere to a hotter one,
and the velocity does not depend on the temperature of the middle sphere.
Moreover, we demonstrated  that the average velocity is determined by the net heat flow 
between the first and the third spheres (see Sec.~\ref{sec:discussion} later)~\cite{Hosaka17}.
This result is consistent with the theoretical framework of ``stochastic 
energetics''~\cite{Sekimoto97,Sekimoto98,SekimotoBook}.
However, a more detailed analysis based on non-equilibrium statistical mechanics is required 
in order to clarify the physical mechanism for the locomotion of a thermally driven micromachine.

It is well-known that systems in thermodynamic equilibrium obey detailed balance meaning 
that transition rates between any two microscopic states are pairwise 
balanced~\cite{KampenBook}.
For non-equilibrium steady state situations, however, detailed balance is broken and 
a probability flux loop should exist in a configuration phase 
space~\cite{Weiss03,Weiss07,Ghanta17,Gnesotto18}.
Such a probability flux has been experimentally measured in the periodic beating of a 
flagellum from \textit{Chlamydomonas reinhardtii} and in the non-periodic fluctuations 
of primary cilia of epithelial cells~\cite{Battle16}.
A similar analysis was performed for non-equilibrium shape fluctuations of semi-flexible 
filaments in a viscoelastic environment~\cite{Gladrow16,Gladrow17}.
Since the existence of a probability flux loop is a direct verification of a non-equilibrium 
steady state, it is a useful concept to characterize driven systems.

In this paper, we discuss the non-equilibrium statistical mechanics of a thermally driven micromachine 
consisting of three spheres and two harmonic springs~\cite{Hosaka17}.
We obtain the steady state conformational distribution function of such a micromachine and calculate 
its probability flux in the corresponding configuration space. 
The obtained probability flux will be expressed in terms of a frequency matrix to discuss the 
non-equilibrium steady state of a micromachine.
The main purpose of our work is to understand the physical mechanism that underlies the locomotion 
of a thermally driven micromachine within the non-equilibrium statistical mechanics.
To this aim, we shall obtain a relation connecting the eigenvalue of the frequency matrix
to the average velocity of a thermally driven micromachine as shown in Eq.~(\ref{Vsymmetric}).
With this relation, we show explicitly that the concept of Purcell's scallop theorem can be generalized 
for thermally driven micromachines.
Together with the results in Ref.~\cite{Hosaka17}, the present study provides an unified description 
of the locomotion of a stochastic elastic micromachine.

In the next Section, we explain our model of a thermally driven three-sphere micromachine
by introducing the coupled Langevin equations for the two spring extensions.
In Sec.~\ref{sec:distribution}, we determine the steady state probability distribution function 
of an elastic micromachine. 
By employing the Fokker-Planck equation in Sec.~\ref{sec:flux}, we obtain the steady state 
probability flux in the configuration space.
From the Gaussian probability flux, we calculate the frequency matrix and the flux rotor. 
In Sec.~\ref{sec:velocity}, the average velocity of a micromachine will be expressed in terms 
of the obtained quantities characterizing the scale of non-equilibrium.
Finally, a summary of our work and some discussion are given in Sec.~\ref{sec:discussion}. 
In Appendix, we give a matrix representation of linear stochastic dynamical systems, 
which is useful to understand our results from a general point of view.

\section{Thermally driven three-sphere micromachine}
\label{sec:elastic}

We first explain the model of a thermally driven elastic micromachine that was introduced 
before by the present authors~\cite{Hosaka17}. 
As schematically shown in Fig.~\ref{model}, this model consists of three hard spheres 
connected by two harmonic springs.  
For the sake of simplicity, we assume that the two springs are identical, and the common
spring constant and the natural length are given by $K$ and $\ell$, respectively.
Then the total elastic energy is given by 
\begin{align}
E = \frac{K}{2}(x_2 - x_1 - \ell)^2 + \frac{K}{2}(x_3 - x_2 - \ell)^2,
\end{align}
where $x_i(t)$ ($i=1, 2, 3$) are the positions of the three spheres in a one-dimensional 
coordinate system, and we also assume $x_1<x_2<x_3$ without loss of generality.
In our previous model for an elastic swimmer~\cite{Yasuda17,Kuroda19}, the natural length 
of each spring was assumed to undergo a prescribed cyclic motion in time, $\ell(t)$, representing 
internal states of the micromachine.
However, for a thermally driven micromachine~\cite{Hosaka17}, as we discuss in this paper,
$\ell$ is taken to be a constant.

We consider a situation in which the three spheres are in thermal equilibrium 
with independent heat baths at temperatures $T_i$  ($i=1, 2, 3$)~\cite{Hosaka17}.
When these temperatures are different, the system is inevitably driven out of equilibrium 
because heat flux from a hotter sphere to a colder one is generated. 
The equations of motion of the three spheres are written in the form of Langevin 
equations as 
\begin{align}
\dot{x}_1 &= \frac{K}{\zeta_1}(x_2-x_1-\ell)+\left(\frac{2T_1}{\zeta_1}\right)^{1/2}\xi_1,
\label{x1dot} \\
\dot{x}_2 &= - \frac{K}{\zeta_2}(x_2-x_1-\ell)+\frac{K}{\zeta_2}(x_3-x_2-\ell)
\nonumber \\
& +\left( \frac{2T_2}{\zeta_2} \right)^{1/2} \xi_2, 
\label{x2dot} \\
\dot{x}_3 &=- \frac{K}{\zeta_3}(x_3-x_2-\ell)+\left( \frac{2T_3}{\zeta_3} \right)^{1/2} \xi_3,
\label{x3dot}
\end{align}
where dot indicates the time derivative, $\zeta_i$ ($i=1, 2, 3$) is the friction coefficient for 
$i$-th sphere, and the Boltzmann constant is set to unity hereafter 
[except in Eqs.~(\ref{Vgamma2}) and (\ref{Vs0})].
Furthermore, $\xi_i(t)$ ($i=1, 2, 3$) is a zero mean and unit variance Gaussian white noise, 
independent for all the spheres: 
\begin{align}
& \langle \xi_i(t)\rangle  =0,
\\
& \langle \xi_i(t)\xi_j(t') \rangle  = \delta_{ij}\delta{(t-t')}.
\end{align}
In contrast to Ref.~\cite{Hosaka17}, we do not consider hydrodynamic interactions 
acting between different spheres in Eqs.~(\ref{x1dot})--(\ref{x3dot}).
Hence, the locomotion of a micromachine is not explicitly taken into account in this paper, 
but such a treatment is sufficient for the statistical analysis of the configurational properties.
Corrections due to hydrodynamic interactions will briefly be discussed in Sec.~\ref{sec:discussion}.

In the following argument, it is convenient to introduce the two spring extensions with 
respect to $\ell$:
\begin{align}
r_{12}=x_2-x_1-\ell,~~~~~r_{23}=x_3-x_2-\ell.
\label{r12r23}
\end{align}
From Eqs.~(\ref{x1dot})--(\ref{x3dot}), we obtain the Langevin equations for $r_{12}(t)$ and 
$r_{23}(t)$ as~\cite{Grosberg15,Netz18}
\begin{align}
\dot{r}_{12} &= -\frac{K}{\zeta_{12}} r_{12} + \frac{K}{\zeta_2} r_{23} + 
\left( \frac{2T_{12}}{\zeta_{12}} \right)^{1/2} \xi_{12}, 
\label{eqr12} \\
\dot{r}_{23} &=  \frac{K}{\zeta_2}r_{12} - \frac{K}{\zeta_{23}} r_{23} +
\left( \frac{2T_{23}}{\zeta_{23}} \right)^{1/2} \xi_{23}.
\label{eqr23}
\end{align}
Here we have introduced the relevant effective friction coefficient 
\begin{align}
\zeta_{ij} = \frac{\zeta_i\zeta_j}{\zeta_i + \zeta_j},
\label{zetaij}
\end{align}
and the mobility-weighted average temperature
\begin{align}
T_{ij} = \frac{\zeta_j T_i + \zeta_i T_j}{\zeta_i + \zeta_j}. 
\label{Tij}
\end{align}
The average temperatures $T_{12}$ and $T_{23}$ can be regarded as two effective 
temperatures characterizing non-equilibrium behaviors of a thermally driven 
micromachine in the reduced configuration space, i.e., in the $(r_{12},r_{23})$ space.

The definition of the effective temperature $T_{ij}$ arises from the condition that the newly 
defined noises $\xi_{12}(t)$ and $\xi_{23}(t)$ in Eqs.~(\ref{eqr12}) and (\ref{eqr23}), 
respectively, satisfy the following statistical properties:
\begin{align}
& \langle \xi_{12}(t)\rangle= \langle \xi_{23}(t)\rangle = 0,
\\
& \langle \xi_{12}(t)\xi_{12}(t')\rangle=  \delta(t-t'), 
\\
& \langle \xi_{23}(t)\xi_{23}(t')\rangle=  \delta(t-t'), 
\\
& \langle \xi_{12}(t)\xi_{23}(t')\rangle= 
- \frac{T_{2}}{\zeta_{2}}\left( \frac{\zeta_{12}\zeta_{23}}{T_{12}T_{23}}\right)^{1/2} \delta(t - t').
\label{xi12xi23}
\end{align}
It is important to note that the strength of the cross-correlation  $\langle \xi_{12}(t)\xi_{23}(t')\rangle$ 
in Eq.~(\ref{xi12xi23}) is negative and its amplitude differs from unity.
This noise property turns out to be important when we discuss the Fokker-Planck equation 
for the probability distribution function in Sec.~\ref{sec:flux}.

\section{Steady state distribution function}
\label{sec:distribution}

\subsection{Covariance matrix}

In this Section, we shall investigate the conformational distribution of $r_{12}$ and 
$r_{23}$ that obey the coupled Langevin equations given by Eqs.~(\ref{eqr12}) and (\ref{eqr23}).
For this purpose, we introduce a Fourier representation of any function $f(t)$ as  
\begin{align}
f(t) = \int_{-\infty}^{\infty} \frac{d\omega}{2\pi} \, f[\omega] e^{i \omega t},
\end{align}
where $\omega$ is the frequency and $f[\omega]$ is the Fourier component.
We rewrite Eqs.~(\ref{eqr12}) and (\ref{eqr23}) in terms of $r_{12}[\omega]$ and 
$r_{23}[\omega]$ and solve for them.
After some calculations, they become 
\begin{align}
r_{12}[\omega] & =  - \frac{1}{b} \biggl[ 
\left(\frac{i\omega \zeta_2}{K} + \frac{\zeta_2}{\zeta_{23}} \right) 
\left(\frac{2 T_{12}}{\zeta_{12}}\right)^{1/2}\xi_{12}[\omega]  
\nonumber \\
& + \left(\frac{2T_{23}}{\zeta_{23}}\right)^{1/2} \xi_{23}[\omega] \biggr], 
\label{r12omega} \\
r_{23}[\omega] & = - \frac{1}{b} \biggl[ \left(
\frac{ i\omega \zeta_2}{K} + \frac{\zeta_2}{\zeta_{12}} \right)
\left(\frac{2 T_{23}}{\zeta_{23}}\right)^{1/2}\xi_{23}[\omega] 
\nonumber \\
& + \left(\frac{2T_{12}}{\zeta_{12}}\right)^{1/2} \xi_{12}[\omega] \biggr], 
\label{r23omega}
\end{align}
where the common quantity $b$ in the denominators is 
\begin{align}
b= \frac{\zeta_2}{K} \omega^2 
- i \zeta_2\left( \frac{1}{\zeta_{12}}+\frac{1}{\zeta_{23}}\right) \omega 
+ \frac{K}{\zeta_2}\left( 1 - \frac{\zeta_2^2}{\zeta_{12}\zeta_{23}} \right).
\end{align}

Using Eqs.~(\ref{r12omega}) and (\ref{r23omega}), one can calculate the three  
correlation functions 
$\langle r_{12}[\omega] r_{12}[\omega'] \rangle$,
$\langle r_{23}[\omega] r_{23}[\omega'] \rangle$, and
$\langle r_{12}[\omega] r_{23}[\omega'] \rangle$
in the frequency domain.
Then the two variances $\sigma_{12}^2$, $\sigma_{23}^2$ and the covariance $\sigma_{13}$
(or the equal-time correlation functions) are obtained as  
\begin{align}
\sigma_{12}^2  & = \int_{-\infty}^{\infty}\frac{d\omega}{2\pi}
\int_{-\infty}^{\infty}\frac{d\omega'}{2\pi} \,
\langle r_{12}[\omega] r_{12}[\omega'] \rangle
\nonumber \\
& = \frac{1}{K}(T_{12} + \zeta_{12}\Delta), 
\label{sigma12}
\end{align}
\begin{align}
\sigma_{23}^2 & = \int_{-\infty}^{\infty}\frac{d\omega}{2\pi}
\int_{-\infty}^{\infty}\frac{d\omega'}{2\pi} \,
\langle r_{23}[\omega] r_{23}[\omega'] \rangle
\nonumber \\
& = \frac{1}{K}(T_{23} + \zeta_{23}\Delta),
\label{sigma23}
\end{align}
\begin{align}
\sigma_{13} & = \int_{-\infty}^{\infty}\frac{d\omega}{2\pi}
\int_{-\infty}^{\infty}\frac{d\omega'}{2\pi} \,
\langle r_{12}[\omega] r_{23}[\omega'] \rangle
\nonumber \\
& = \frac{\zeta_2  \Delta}{K},
\label{sigma13}
\end{align}
where we have used the notation
\begin{align}
\Delta = \frac{\zeta_{12} \zeta_{23}(T_{12} + T_{23}-2T_2)}
{(\zeta_{12} + \zeta_{23})(\zeta_2^2 -\zeta_{12} \zeta_{23})}.
\end{align}

\begin{figure}[tbh]
\begin{center}
\includegraphics[scale=0.3]{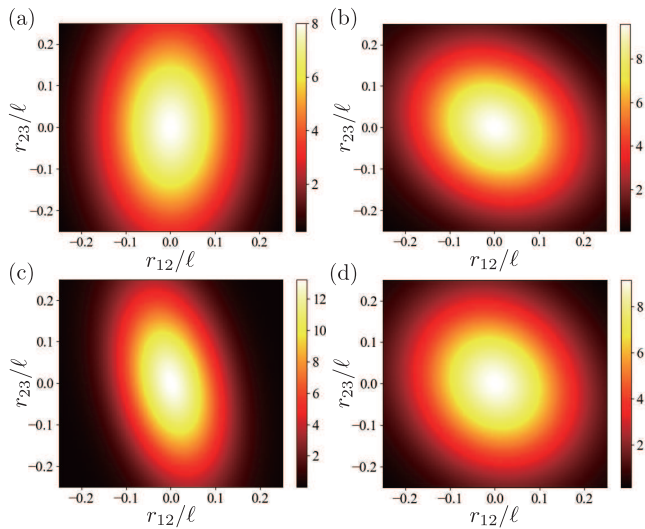}
\end{center}
\caption{(Color online)　
Dimensionless probability distribution function $p({\bm r}) \ell^2$
given by Eq.~(\ref{distribution2}) as a function of $r_{12}/\ell$ and $r_{23}/\ell$.
The parameters are 
(a) $\zeta_1=\zeta_2 = \zeta_3$,
$\tau_1 = 1/900$, $\tau_2 = 41/900$, and  $\tau_3 = 81/900$
(these temperatures satisfy $\tau_1+\tau_3=2\tau_2$),
(b) $\zeta_1=\zeta_2 = \zeta_3$,
$\tau_1 = \tau_3 = 25/900$, and  $\tau_2 = 41/900$, 
(c) $\zeta_1/\zeta_2 = 0.5$, $\zeta_3/\zeta_2 = 5$,
$\tau_1 = 1/900$, $\tau_2 = 41/900$, and  $\tau_3 = 81/900$
(these temperatures are the same as in (a)),
(d) $\zeta_1/\zeta_2 = 0.5$, $\zeta_3/\zeta_2 = 5$,
$\tau_1 = \tau_3 = 25/900$, and  $\tau_2 = 41/900$
(these temperatures are the same as in (b)).
}
\label{distribution}
\end{figure}

With these results, we construct a symmetric covariance matrix $\mathbf C$ defined by 
\begin{align}
{\mathbf C} & = 
\begin{pmatrix} 
\sigma_{12}^2 &  \sigma_{13} 
\\
\sigma_{13}  & \sigma_{23}^2
\end{pmatrix}. 
\label{covarianceC}
\end{align}
Then the inverse of the covariance matrix can be simply given by  
\begin{align}
\mathbf C^{-1} =
\frac{1}{\sigma_{12}^2 \sigma_{23}^2 (1-\rho^2)} 
\begin{pmatrix} 
\sigma_{23}^2 &  - \sigma_{13} 
\\
- \sigma_{13}  & \sigma_{12}^2
\end{pmatrix},
\label{inversecovariance} 
\end{align}
where the correlation factor $\rho$ is defined by  
\begin{align}
\rho = \frac{\sigma_{13}}{\sigma_{12} \sigma_{23}} = 
\frac{\zeta_2 \Delta}{[(T_{12} + \zeta_{12}\Delta)(T_{23}+\zeta_{23}\Delta)]^{1/2}}.
\label{correlation}
\end{align}
Notice that the absolute value of the correlation factor should satisfy the condition 
$\vert \rho \vert < 1$.
As explained in Ref.~\cite{Weiss03} and summarized in Appendix, the steady state covariance 
matrix can generally be obtained by solving the corresponding Lyapunov equation 
[see Eq.~(\ref{lyapunov})].

When the three friction coefficients are all identical, i.e., $\zeta_1=\zeta_2=\zeta_3$, 
the above correlation factor reduces to 
\begin{align}
\rho = \frac{2 ( T_1 + T_3 - 2T_2)}
{ [( 7T_1 + 4T_2 + T_3)( T_1 + 4 T_2 + 7T_3 ) ]^{1/2}}.
\label{symcorr}
\end{align}
Here, we see that $\rho$ generally vanishes when $T_1 + T_3 = 2T_2$, i.e, the temperature
of the middle sphere coincides with the average temperature between the first and the 
third spheres.
Obviously, $\rho$ vanishes in thermal equilibrium, $T_1=T_2=T_3$, although $\rho=0$ 
does not mean that the micromachine is in thermal equilibrium.

\subsection{Distribution function}

Next we consider the steady state probability distribution function $p({\bm r})$, where 
${\bm r}=(r_{12}, r_{23})^{\mathsf T}$ and ``${\mathsf T}$" indicates the transpose.
Owing to the reproductive property of Gaussian distributions~\cite{KampenBook}, 
$p({\bm r})$ should also be a Gaussian function for the present linear problem and is given by 
\begin{align}
p({\bm r}) = {\cal N} \exp \left[ -  \frac{1}{2} {\bm r}^{\mathsf T} {\mathbf C}^{-1} 
{\bm r} \right],
\label{gaussian}
\end{align}
where ${\cal N}$ is the normalization factor. 
Using Eq.~(\ref{inversecovariance}) for ${\mathbf C}^{-1}$, we can write the explicit form 
of the distribution function as 
\begin{align}
& p({\bm r}) = \frac{1}{ 2\pi \sigma_{12} \sigma_{23} (1 - \rho^2)^{1/2}} 
\nonumber \\
& \times \exp \biggl[ -\frac{1}{2(1-\rho^2)} 
\biggl( \frac{r_{12}^2}{\sigma_{12}^2} 
- 2\rho\frac{r_{12} r_{23} }{\sigma_{12} \sigma_{23}}  
+ \frac{r_{23}^2}{\sigma_{23}^2}\biggr) \biggr],
\label{distribution2}
\end{align}
together with Eqs.~(\ref{sigma12}), (\ref{sigma23}), and (\ref{correlation}).
Despite the fact that the micromachine is out of equilibrium, the extensions of the two springs 
obey the Boltzmann-type distribution that is characterized by the effective temperatures 
$T_{12}$ and $T_{23}$~\cite{Grosberg15}.
Our further analysis crucially depends on the result of Eq.~(\ref{distribution2}).

Let us introduce the following dimensionless temperature parameter 
\begin{align}
\tau_i = \frac{2T_i}{K\ell^2},
\end{align}
which is the ratio between the thermal energy of each sphere and the spring elastic energy
(recall $k_{\rm B}=1$).
In Fig.~\ref{distribution}, we plot the dimensionless steady state distribution function $p({\bm r})$ 
in Eq.~(\ref{distribution2}) as a function of $r_{12}/\ell$ and $r_{23}/\ell$ for different parameter
combinations.
The absolute value of the friction coefficient is not required for these plots because we are 
focusing only on the stationary state distribution.
The parameters in Fig.~\ref{distribution}(a) are 
$\zeta_1=\zeta_2 = \zeta_3$, 
$\tau_1 = 1/900$, $\tau_2 = 41/900$, and  $\tau_3 = 81/900$.
Notice that these temperatures satisfy $\tau_1+\tau_3=2\tau_2$ and hence $\rho=0$
according to Eq.~(\ref{symcorr}).
This means that the distribution function $p({\bm r})$ is simply a product of two Gaussian functions,
and there is no correlation between $r_{12}$ and $r_{23}$ for these temperatures.
The distribution function is elongated in the $r_{23}$-direction because $\tau_1<\tau_3$.

The parameters in Fig.~\ref{distribution}(b) are $\zeta_1= \zeta_2 = \zeta_3$,
$\tau_1 = \tau_3 = 25/900$, and  $\tau_2 = 41/900$. 
Here the temperatures of the first and the third spheres are identical although 
the correlation factor obtained from Eq.~(\ref{symcorr}) is negative, $\rho<0$.
The correlation between $r_{12}$ and $r_{23}$ causes a tilt of the elongated distribution and 
the tilt direction has a negative slope.
Here the whole plot is symmetric with respect to the straight line $r_{12}=r_{23}$.
In spite of the correlation between $r_{12}$ and $r_{23}$, the system is thermally balanced 
because the effective temperatures $T_{12}$ and $T_{23}$ are identical.
As we mention in the next Section, the micromachine behaves as if it were in thermal equilibrium
in the reduced configuration space.

Keeping the temperature parameters $\tau_i$ the same as in Fig.~\ref{distribution}(a) and (b), 
we introduce asymmetries in the friction coefficients, e.g., $\zeta_1/\zeta_2 = 0.5$ and 
$\zeta_3/\zeta_2 = 5$, in Fig.~\ref{distribution}(c) and (d), respectively.
Then the correlation factor should be estimated according to Eq.~(\ref{correlation}).
In Fig.~\ref{distribution}(c), there is a negative correlation between $r_{12}$ and $r_{23}$ owing 
to the different friction coefficients.
On the other hand, Fig.~\ref{distribution}(d) is no longer symmetric with respect to the line 
$r_{12}=r_{23}$.
Since $T_{12} \neq T_{23}$ in this case [see Eq.~(\ref{Tij})], the system is not thermally balanced
and exhibits non-equilibrium behaviors.

So far, we have calculated the steady state conformational distribution 
function $p$
of a thermally driven micromachine. 
We have shown that it is given by a Gaussian function characterized by the covariance 
matrix ${\mathbf C}$.
In the next Section, we shall calculate the probability flux by using the obtained 
distribution function.

\section{Probability flux and related quantities}
\label{sec:flux}

\subsection{Probability flux}

Next, we calculate the probability flux that can be used to characterize the non-equilibrium 
properties of a three-sphere micromachine. 
The Fokker-Planck equation corresponding to Eqs.~(\ref{eqr12}) and (\ref{eqr23}) can be 
written for the time-dependent conformational probability distribution function $p({\bm r},t)$ 
as~\cite{RiskenBook,ZwanzigBook} 
\begin{align}
\dot{p} = - \nabla \cdot {\bm j},
\label{FPeq}
\end{align}
expressing the fact that probability is neither created nor annihilated.
In the above, the two-dimensional nabla operator indicates 
$\nabla=(\partial/\partial r_{12}, \partial/\partial r_{23})^{\mathsf T}$ in the configuration space
and the vector ${\bm j}= (j_{12}, j_{23})^{\mathsf T}$ is the two-dimensional probability flux 
given by 
\begin{align}
{\bm j} = {\mathbf A} {\bm r} p - {\mathbf D} \nabla p,
\label{probflux}
\end{align}
where the matrices ${\mathbf A}$ and ${\mathbf D}$ are 
\begin{align}
{\mathbf A} = 
\begin{pmatrix} 
-K/\zeta_{12} &  K/\zeta_2 
\\
K/\zeta_2 & - K/\zeta_{23}
\end{pmatrix}, 
\label{Amatrix}
\end{align}
and
\begin{align}
{\mathbf D}  =
\begin{pmatrix} 
T_{12}/\zeta_{12} &  -T_2/\zeta_2
\\
 -T_2/\zeta_2 & T_{23}/\zeta_{23}
\end{pmatrix}, 
\label{Dmatrix}
\end{align}
respectively.
Here ${\mathbf A}$ describes the linear deterministic dynamics (without noise) in Eqs.~(\ref{eqr12}) and 
(\ref{eqr23}), whereas ${\mathbf D}$ is called the diffusion matrix (see also Appendix). 
The non-zero off-diagonal components of ${\mathbf D}$ originate from the noise correlation 
shown in Eq.~(\ref{xi12xi23}).

In a steady state, i.e., $\dot{p}= 0$, the probability flux ${\bm j}({\bm r})$ can be obtained from 
Eq.~(\ref{probflux}) by using the Gaussian probability distribution function $p$ in Eq.~(\ref{distribution2}):
\begin{align}
j_{12}({\bm r}) &= \biggl[ K \biggl( \frac{r_{23}}{\zeta_2} - \frac{r_{12}}{\zeta_{12}} \biggr) 
+ \frac{1}{(1 - \rho^2)} \biggl\{ \frac{T_{12}r_{12}}{\zeta_{12} \sigma_{12}^2} 
\nonumber  \\
& + \frac{\rho}{\sigma_{12} \sigma_{23}}\biggl( \frac{T_2 r_{12}}{\zeta_2} 
- \frac{T_{12} r_{23}}{\zeta_{12}} \biggr) -\frac{T_2 r_{23}}{\zeta_{2} \sigma_{23}^2} \biggr\} \biggr] p, 
\label{J12}
\end{align}
\begin{align}
j_{23}({\bm r}) &= \biggl[ K \biggl(  \frac{r_{12}}{\zeta_{2}} - \frac{r_{23}}{\zeta_{23}} \biggr) + 
\frac{1}{(1 - \rho^2)} \biggl\{ -\frac{T_2 r_{12}}{\zeta_2 \sigma_{12}^2}  
\nonumber \\
& + \frac{\rho}{\sigma_{12} \sigma_{23}}\biggl( \frac{T_2 r_{23}}{\zeta_2} 
- \frac{T_{23} r_{12}}{\zeta_{23}} \biggr) +\frac{T_{23} r_{23}}{\zeta_{23} \sigma_{23}^2} 
\biggr\} \biggr] p.
\label{J23}
\end{align}
One can confirm that this probability flux is divergence-free, i.e., 
$\nabla \cdot {\bm j}=0$, so that the steady state condition is satisfied.

\subsection{Frequency matrix}

The above probability flux ${\bm j}$ can be conveniently expressed in terms of a 
frequency matrix ${\bm \Omega}$ as~\cite{Weiss03}
\begin{align}
{\bm j} ={\bm \Omega} {\bm r}p, 
\label{jomega}
\end{align}
where the explicit expression of ${\bm \Omega}$ is given by
\begin{widetext}
\begin{align}
{\bm \Omega} = 
\begin{pmatrix}
\displaystyle
-\frac{K}{\zeta_{12}} + \frac{1}{(1 - \rho^{2})}
\left( \frac{T_{12}}{\zeta_{12}{\sigma_{12}}^{2}} + \frac{\rho T_{2}}{\sigma_{12} \sigma_{23} \zeta_{2}} \right)
& 
\displaystyle
\frac{K}{\zeta_{2}} - \frac{1}{(1 - \rho^{2})}
\left( \frac{T_{2}}{\zeta_{2}{\sigma_{23}}^{2}} + \frac{\rho T_{12}}{\sigma_{12} \sigma_{23} \zeta_{12}} \right)
\\[2.0ex]
\displaystyle
\frac{K}{\zeta_{2}} - \frac{1}{(1 - \rho^{2})}
\left( \frac{T_{2}}{\zeta_{2}{\sigma_{12}}^{2}} + \frac{\rho T_{23}}{\sigma_{12} \sigma_{23} \zeta_{23}} \right)
 & 
 \displaystyle
-\frac{K}{\zeta_{23}} + \frac{1}{(1 - \rho^{2})}
\left( \frac{T_{23}}{\zeta_{23}{\sigma_{23}}^{2}} + \frac{\rho T_{2}}{\sigma_{12} \sigma_{23} \zeta_{2}} \right) 
\end{pmatrix}.
\label{genfrequencymatrix}
\end{align}
\end{widetext}
Since ${\bm j}$ is divergence-free, ${\bm \Omega}$ should be a traceless matrix~\cite{Weiss07}.
Using Eqs.~(\ref{zetaij}) and (\ref{Tij}), we indeed find that ${\bm \Omega}$ is traceless.
In Appendix, we generally show that the frequency matrix ${\bm \Omega}$ can be expressed 
as ${\bm \Omega}={\mathbf A}+ {\mathbf D} {\mathbf C}^{-1}$ within the matrix formulation
[see Eq.~(\ref{Omega})].

\begin{figure}[tbh]
\begin{center}
\includegraphics[scale=0.3]{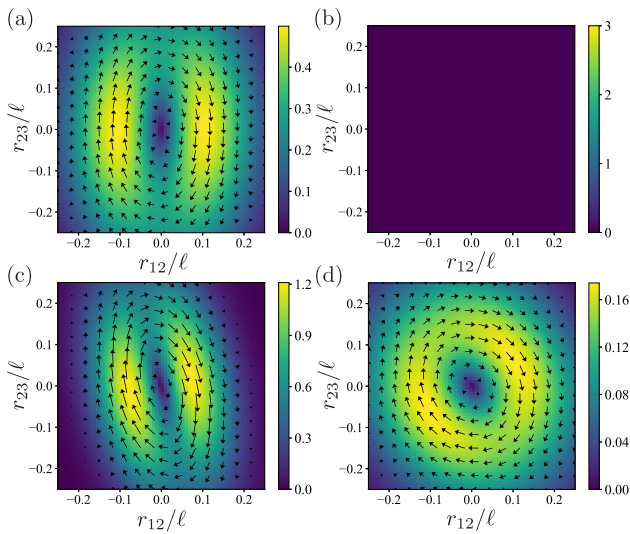}
\end{center}
\caption{(Color online)　
Dimensionless probability flux vector ${\bm j} ({\bm r}) \zeta_2 \ell/K$ 
given by Eq.~(\ref{jomega}) as a function of $r_{12}/\ell$ and $r_{23}/\ell$.
The parameter values are the same as in Fig.~(\ref{distribution}).
There is no probability flux in (b).
The color scale indicates the magnitude of the flux vector ${\bm j}$.}
\label{flux}
\end{figure}

When the friction coefficients are identical, $\zeta_1=\zeta_2 = \zeta_3=\zeta$, the frequency matrix 
is simplified to  
\begin{align}
{\bm \Omega} & = \frac{K(T_1 - T_3)}{\zeta c}
\nonumber \\ 
& \times
\begin{pmatrix}
2 (T_1+ T_3 -2 T_2) & - (7 T_1 + 4 T_2 + T_3) 
\\
 T_1 + 4 T_2 + 7 T_3&  - 2 (T_1+ T_3 -2 T_2)
\end{pmatrix},
\label{frequencymatrix}
\end{align}
where the quantity $c$ in the denominators is
\begin{align}
c = T_1^2+16 T_1 T_2+14 T_1 T_3+16 T_2 T_3+T_3^2.
\label{quantityc}
\end{align}
Here, we explicitly see that ${\bm \Omega}$ is proportional to $T_1 - T_3$, and hence the probability 
flux ${\bm j}$ vanishes when $T_1 = T_3$ for any $T_2$ of the middle sphere.
Indeed, ${\bm j}=0$ or ${\bm \Omega}=0$ is a necessary and sufficient condition for physical 
situations in which detailed balance is satisfied~\cite{Weiss03}.
When the system is in thermal equilibrium and detailed balance holds, the probability of a 
transition between any two states is the same as the probability of the reverse transition.
As a result, a net probability flux does not exist.
In Appendix, we generally show that the necessary and sufficient condition for detailed balance 
is given by the commutation relation ${\mathbf A}{\mathbf D} - {\mathbf D}{\mathbf A}^{\mathsf T} =0$
[see Eq.~(\ref{detailed})].

\begin{figure}[tbh]
\begin{center}
\includegraphics[scale=0.3]{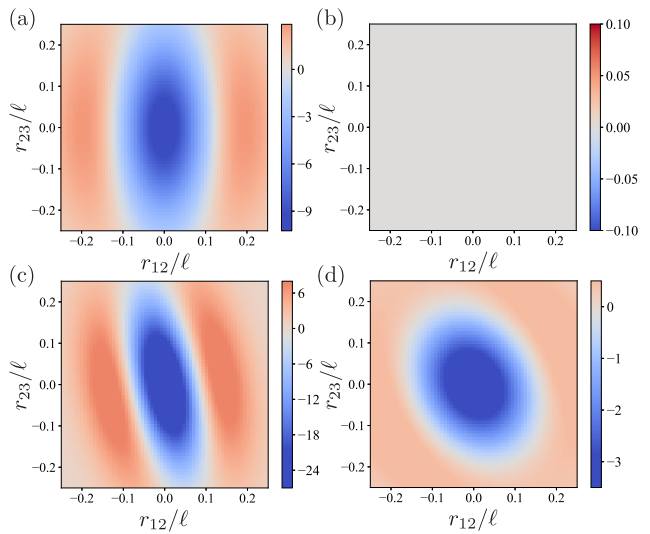}
\end{center}
\caption{(Color online)　
Dimensionless flux rotor $s ({\bm r}) \zeta_2 \ell^2/K$ given by Eq.~(\ref{sr}) as a 
function of $r_{12}/\ell$ and $r_{23}/\ell$.
The parameter values are the same as in Fig.~(\ref{distribution}).
There is no flux rotor in (b).
}
\label{rotation}
\end{figure}

When $T_1 \neq T_3$, on the other hand, we have non-zero frequency matrix ${\bm \Omega}$.
Then detailed balance is violated and the micromachine is in a non-equilibrium steady state.
Hence the frequency matrix is an important measure to quantify the out of equilibrium 
properties of a thermally driven micromachine.
When $T_1 \neq T_3$, the two eigenvalues of ${\bm \Omega}$ in Eq.~(\ref{frequencymatrix})
are given by 
\begin{align}
\gamma = \pm \frac{i \sqrt{3} K(T_1 - T_3)}{\zeta c^{1/2}}.
\label{eigenvalues}
\end{align}
Since these eigenvalues are purely imaginary, the probability current in the configuration space is 
rotational~\cite{Weiss03, Gnesotto18}.
In the above, the essential frequency scale is set by $\gamma \sim K/\zeta$ which characterizes 
the speed of the rotational motion.

In Fig.~\ref{flux}, we plot the dimensionless probability flux vector ${\bm j}({\bm r})$ in 
Eq.~(\ref{jomega}) as a function of $r_{12}/\ell$ and $r_{23}/\ell$ 
by using the same parameter values as in Fig.~\ref{distribution}.
Note that the color scale indicates the magnitude of the flux vector. 
In Fig.~\ref{flux}(a), (c), and (d), clockwise flux loops can be clearly seen.
When the friction coefficients are identical as in Fig.~\ref{flux}(a), the existence of a flux 
loop is consistent with a directed motion of a three-sphere micromachine~\cite{Hosaka17}, 
as we further discuss in Sec.~\ref{sec:velocity}.

In Fig.~\ref{flux}(b) with $\tau_1=\tau_3$, however, the probability flux does not exist 
simply because ${\bm \Omega}=0$.
Asymmetries both in the temperatures and the friction coefficients lead to a flux loop 
in Fig.~\ref{flux}(c).
The presence of a flux loop in Fig.~\ref{flux}(d) even for $\tau_1=\tau_3$ is due to the 
asymmetry in the friction coefficients.
Although the case with different friction coefficients was not studied in Ref.~\cite{Hosaka17}, 
we expect that a finite flux loop leads to a directed motion of a three-sphere 
micromachine despite the balanced temperature distribution.

\subsection{Flux rotor}

To further characterize the strength of the probability flux loop in terms of a scalar quantity, we 
consider the following flux rotor~\cite{Dotsenko13}:
\begin{align}
s({\bm r}) & = \frac{\partial}{\partial r_{12}} j_{23} - \frac{\partial}{\partial r_{23}} j_{12}
\nonumber \\
& =\frac{p(w_0 + w_{12}r_{12}^2 + w_{23} r_{23}^2 +w_{13} r_{12} r_{23})}
{\zeta_2 \zeta_{12} \zeta_{23} \sigma_{12}^4 \sigma_{23}^4 (1 - \rho^2)^2},
\label{sr}
\end{align}
where $w_0$, $w_{12}$, $w_{23}$, and $w_{13}$ are 
\begin{align}
w_0 & = \sigma_{12}^2 \sigma_{23}^2 (1 - \rho^{2}) 
[\zeta_{12} \zeta_{23} T_2 (\sigma_{12}^2 - \sigma_{23}^2) 
\nonumber \\
& + \zeta_2 \sigma_{12} \sigma_{23} \rho (\zeta_{23} T_{12} -\zeta_{12} T_{23})],
\label{w0}
\\
w_{12} &= [\zeta_2 \sigma_{12} \sigma_{23} \rho  (\zeta_{12} T_{23} - \zeta_{23} T_{12}) 
\nonumber \\
& - K \zeta_{23} \sigma_{12}^2 \sigma_{23} (1 - \rho^{2})
(\zeta_{12} \sigma_{23} - \zeta_2 \sigma_{12} \rho ) 
\nonumber \\
& + \zeta_{12} \zeta_{23} T_2 (\sigma_{23}^2 - \rho^2 \sigma_{12}^2)] \sigma_{23}^2,
\\
w_{23} &= [\zeta_2 \sigma_{12} \sigma_{23} \rho (\zeta_{12} T_{23} - \zeta_{23} T_{12}) 
\nonumber \\
& - K \zeta_{12} \sigma_{12} \sigma_{23}^2 (1 - \rho^2)
( \zeta_2 \sigma_{23} \rho -\zeta_{23} \sigma_{12})
\nonumber \\
& - \zeta_{12} \zeta_{23} T_2 ( \sigma_{12}^2 - \rho^2 \sigma_{23}^2 )] \sigma_{12}^2,
\\
w_{13} &= [2 \zeta_{12} \zeta_{23} T_2 \rho  (\sigma_{12}^2 - \sigma_{23}^2) 
\nonumber \\
& + K \zeta_2 \sigma_{12} \sigma_{23} (1- \rho^2) (\zeta_{12} \sigma_{23}^2 
-\zeta_{23} \sigma_{12}^2) 
\nonumber \\
& + \zeta_2 \sigma_{12} \sigma_{23} (1 + \rho^2) (\zeta_{23} T_{12} - \zeta_{12} T_{23})]
\sigma_{12}\sigma_{23}.
\end{align}

In Fig.~\ref{rotation}, we plot the dimensionless flux rotor $s({\bm r})$ in Eq.~(\ref{sr}) as a 
function of $r_{12}/\ell$ and $r_{23}/\ell$ by using the same parameter values as in Fig.~\ref{distribution}.
In Fig.~\ref{rotation}(a), (c), and (d), the flux rotor $s$ takes negative values around 
the origin of the configuration space, ${\bm r}=0$, whereas it takes positive values in the outer regions.
In Fig.~\ref{rotation}(a), the distribution of $s$ is elongated in the $r_{23}$-direction, while there 
is no correlation between $r_{12}$ and $r_{23}$.
The flux rotor vanishes in Fig.~\ref{rotation}(b) simply because ${\bm \Omega}=0$ for 
$\tau_1=\tau_3$.
When the friction coefficients are different as in Fig.~\ref{rotation}(c) and (d), the flux rotor 
exhibits a negative correlation.

\begin{figure}[tbh]
\begin{center}
\includegraphics[scale=0.5]{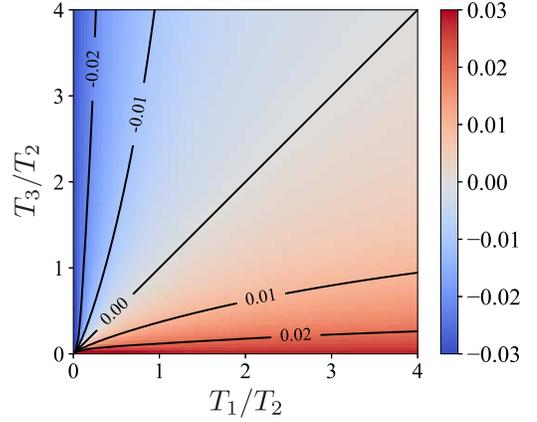}
\end{center}
\caption{(Color online)　
Dimensionless flux rotor $\pi \zeta T_2 s_0/(16 \sqrt{3} K^2)$ given by Eq.~(\ref{s0same})
as a function of $T_1/T_2$ and $T_3/T_2$.
The flux rotor is evaluated at ${\bm r}=0$ when the friction coefficients are identical.
Different curves are the contour lines for the corresponding values.}
\label{rotors0}
\end{figure}

As considered in Ref.~\cite{Dotsenko13}, one can focus on the strength of the flux rotor at the origin 
of the configuration space, i.e, $s({\bm r}=0)$.
This quantity, denoted as $s_0$, is given by
\begin{align}
& s_0  =\frac{1}{2 \pi  \zeta_2\zeta_{12} \zeta_{23} \sigma_{12}^3 \sigma_{23}^3 (1-\rho ^2)^{3/2}} 
\nonumber \\
& \times \left[\zeta_{12} \zeta_{23}T_{2} (\sigma_{12}^2-\sigma_{23}^2) +\zeta_{2}
\sigma_{12} \sigma_{23} \rho (\zeta_{23}T_{12}- \zeta_{12}T_{23}) \right],
\label{s0}
\end{align}
because only the term proportional to $w_0$ in Eq.~(\ref{sr}) remains.
When the friction coefficients are identical, $\zeta_1=\zeta_2 = \zeta_3=\zeta$, 
Eq.~(\ref{s0}) further reduces to
\begin{align}
s_0= \frac{16 \sqrt{3} K^2(T_1 + T_2 +T_3) (T_1 - T_3)}{\pi  \zeta c^{3/2}},
\label{s0same}
\end{align}
where $c$ is given by Eq.~(\ref{quantityc}).
Here the sign of $s_0$ is purely determined by the temperature difference $T_1-T_3$.
In Fig.~\ref{rotors0}, we plot the dimensionless $s_0$ as a function of $T_1/T_2$ and 
$T_3/T_2$ by using a color representation.

In this Section, starting from the Fokker-Planck equation, we have obtained the steady state probability 
flux ${\bm j}$ that can be expressed in terms of the frequency matrix ${\bm \Omega}$.
The eigenvalues $\gamma$ of the frequency matrix are purely imaginary and proportional to 
the temperature difference.
We have also calculated the flux rotor $s$ as a scalar quantity to characterize the scale of 
non-equilibrium of a micromachine. 
In the next Section, we shall discuss the relation between these quantities and the average 
velocity of a stochastic micromachine.

\section{Average velocity of a micromachine}
\label{sec:velocity}

As mentioned before, the main purpose of this paper is to understand the physical mechanism 
that underlies the locomotion of a thermally driven micromachine within the non-equilibrium 
statistical mechanics.
When the friction coefficients are identical, the average velocity is given by Eq.~(\ref{Vsymmetric})
which is proportional to $T_1-T_3$.
This proportionality appears in several scalar quantities such as the eigenvalues of the 
frequency matrix, $\gamma$ [see Eq.~(\ref{eigenvalues})], or the flux rotor at the origin, 
$s_0$ [see Eq.~(\ref{s0same})].
Hence one can construct linear relations between the average velocity and these scalar quantities.

For example, the absolute value of the average velocity can be expressed in terms of $\gamma$ 
as 
\begin{align}
\vert \langle V \rangle \vert = 
\frac{a}{16\sqrt{3}\ell^2} \frac{c^{1/2}}{K} \vert \gamma'' \vert,
\label{Vgamma1}
\end{align}
where we have used Stokes' law, $\zeta=6 \pi \eta a$, for the friction coefficient of a 
hard sphere, and $\gamma''$ indicates the imaginary part of $\gamma$.
For the sake of completeness, we rewrite the above expression by recovering the Boltzmann 
constant and using Eq.~(\ref{quantityc}) as follows:
\begin{align}
& \vert \langle V \rangle \vert = \frac{a}{16\sqrt{3}\ell^2} 
\nonumber \\
& \times \frac{k_{\rm B} (T_1^2+16 T_1 T_2+14 T_1 T_3+16 T_2 T_3+T_3^2 )^{1/2}}{K} 
\vert \gamma'' \vert.
\label{Vgamma2}
\end{align}
As mentioned before, $\gamma''$ has the dimension of frequency and it determines the 
speed of the rotational motion of the probability flux in the configuration space.

An interesting interpretation of the above expression can be made by comparing it to the 
general form for the average velocity of a three-sphere swimmer.
When the two arms undergo a prescribed deterministic motion, as in the original Najafi-Golestanian 
model~\cite{Golestanian04}, the average velocity is given by~\cite{Golestanian08} 
\begin{align}
\overline{V}=\frac{G}{2} \overline{(r_{12}\dot{r}_{23} - \dot{r}_{12} r_{23})},
\label{GolestanianAjdari}
\end{align}
where $G \sim a/\ell^2$ is a geometric factor that depends only on the structural parameters 
of a swimmer, and the averaging is performed by time integration in a full cycle.
In the above form, the averaging part is proportional to the enclosed area (per unit time)
that is swept in a full cycle in the configuration space.
Equation~(\ref{GolestanianAjdari}) can be regarded as a mathematical representation of 
Purcell's scallop theorem for a three-sphere swimmer in low-Reynolds-number 
fluids~\cite{Purcell77,Lauga11}.

For a thermally driven micromachine, $\langle V \rangle$ is proportional to $a/\ell^2$ as in 
Eq.~(\ref{GolestanianAjdari}).
Moreover, $c^{1/2}$ in Eq.~(\ref{Vgamma1}) can be regarded as  overall thermal energy, 
and $c^{1/2}/K$ essentially represents the explored area in the configuration space by 
random motions of the spheres. 
This interpretation is reasonable because one can easily confirm the relation  
$c^{1/2}/K \sim \sigma_{12} \sigma_{23}$, where  $\sigma_{12}$ and  $\sigma_{23}$
are the variances in Eqs.~(\ref{sigma12}) and (\ref{sigma23}), respectively.

While the time scale in the Najafi-Golestanian model is given by the frequency of the periodic 
arm motions~\cite{Golestanian08}, the corresponding time scale for a thermally driven 
micromachine is set by the eigenvalue $\gamma$ of the frequency matrix [see 
Eq.~(\ref{eigenvalues})].  
Hence Eq.~(\ref{Vgamma1}) or Eq.~(\ref{Vgamma2}) provides us with an essential understanding
concerning the locomotion of a thermally driven micromachine.
It is interesting to note that the main concept of Purcell's scallop theorem can be generalized 
for thermally driven micromachines that undergo random motions rather than deterministic cyclic motions.

In order to take into account the sign of the average velocity, namely, the direction of the 
locomotion, it is convenient to use the flux rotor at the origin $s_0$ [see Eq.~(\ref{s0same})].
After recovering the Boltzmann constant, the average velocity can also be written as 
\begin{align}
& \langle V \rangle = -\frac{\pi a }
{256 \sqrt{3} \ell^2}
\nonumber \\
& \times \frac{k_{\rm B}^2 (T_1^2+16 T_1 T_2+14 T_1 T_3+16 T_2 T_3+T_3^2)^{3/2}}
{K^2 (T_1 + T_2 + T_3)}  s_0.
\label{Vs0}
\end{align}
In the above, the sign of $s_0$ determines the direction of the locomotion.
For example, when $T_1 < T_3$ as in Fig.~\ref{rotation}(a) or in Fig.~\ref{rotors0}, we have $s_0<0$ and 
$\langle V \rangle >0$.
Although Eqs.~(\ref{Vgamma2}) and (\ref{Vs0}) are essentially equivalent, we consider 
that  Eq.~(\ref{Vgamma2}) gives more physical insights to understand the locomotion 
of a thermally driven micromachine.

\section{Summary and discussion}
\label{sec:discussion}

In this paper, we have discussed the non-equilibrium statistical mechanics of a thermally driven 
micromachine made of three spheres and two harmonic springs as previously proposed by the 
present authors~\cite{Hosaka17}. 
First, we have calculated the steady state conformational distribution function of such a 
micromachine and showed that it is given by a Gaussian function characterized by the 
covariance matrix [see Eq.~(\ref{distribution2})].
Using this distribution function, we have obtained the steady state probability flux of a 
micromachine.

The distribution function can be expressed in terms of a traceless frequency matrix [see 
Eqs.~(\ref{genfrequencymatrix}) and (\ref{frequencymatrix})].
When the friction coefficients are all identical, we have shown that the eigenvalues of the 
frequency matrix are proportional to the temperature difference between the first and the third 
spheres  [see Eq.~(\ref{frequencymatrix})].
Moreover, the scale of non-equilibrium of a micromachine can quantitatively be characterized by 
the flux rotor [see Eq.~(\ref{sr}) or Eq.~(\ref{s0})].
As one of the main results of this paper, the average velocity of a thermally driven machine is 
expressed in terms of the eigenvalue of the frequency matrix [see Eq.~(\ref{Vgamma1})].
This expression allows us to generalize the concept of Purcell's scallop theorem that is 
also applicable for thermally driven stochastic micromachines.

An interesting situation to be discussed in more detail is the case when $\zeta_1=\zeta_2=\zeta_3$
and $T_1=T_3 \neq T_2$. 
Microscopically, such a micromachine is in a non-equilibrium steady state because the 
temperature of the second sphere is different from those of the other two spheres.
However, within the configuration space $(r_{12}, r_{23})$, the two relevant average temperatures 
are equal, i.e., $T_{12}=T_{23}$.
Hence the overall state of a micromachine is effectively in thermal equilibrium and detailed balance 
is satisfied, i.e., ${\bm \Omega}=0$.
This is the reason why the average velocity in Eq.~(\ref{Vsymmetric}) vanishes when $T_1=T_3$ 
regardless of the value of $T_2$.

It is known that for some non-equilibrium systems, broken detailed balance doest not need to be 
apparent at larger scales, and they can regain thermodynamic equilibrium when the system 
is coarse grained~\cite{Egolf00,Rupprecht16}.
In our analysis, the reduction of the configuration space has been implicitly assumed because we have focused 
only on the distribution of $r_{12}$ and $r_{23}$, while the center of mass motion of a micromachine
has been neglected.
Within this level of description, an elastic three-sphere micromachine behaves as if it were in thermal 
equilibrium when $T_1=T_3$.

In our previous paper~\cite{Hosaka17}, we argued that the average velocity of a micromachine 
can be related to the ensemble average of heat flows in a steady state. 
According to ``stochastic energetics", the heat gained by $i$-th sphere per unit time is expressed as~\cite{Sekimoto97,Sekimoto98,SekimotoBook}
\begin{align}
\dot{Q}_i=\zeta_i (-\dot x_i+\xi_i) \dot x_i,
\label{dQdt}
\end{align}
where $\dot{x}_i(t)$ and $\xi_i(t)$ are given in Eqs.~(\ref{x1dot})--(\ref{x3dot}).
When the friction coefficients are identical, we showed that the average velocity can also 
be expressed in terms of the lowest-order average heat flows as 
\begin{align}
\langle V \rangle=\frac{a}{8K \ell^2} (\langle \dot{Q}_3 \rangle-
\langle \dot{Q}_1 \rangle ).
\label{vel-heat-sym}
\end{align}
This relation states that the average velocity is determined by the net heat flow between 
the first and the third spheres.
Our result in this paper indicates that the net heat flow between the first and the third spheres 
is also proportional to the eigenvalue of the frequency matrix $\gamma$ or the 
flux rotor $s_0$, as discussed in Sec.~\ref{sec:velocity}.

We mention here that our model of a three-sphere micromachine has a similarity to that 
of two over-damped, tethered spheres coupled by a harmonic spring and also confined between 
two walls~\cite{Gnesotto18,Battle16}. 
Because these two spheres are in contact with heat baths having different temperatures, 
the system can be driven out of equilibrium.
They numerically showed that displacements obey a Gaussian distribution and also found 
probability flux loops that demonstrate the broken detailed balance~\cite{Battle16}.
In Ref.~\cite{Gnesotto18}, an analytical expression of the frequency matrix for this two-sphere 
model was shown to be proportional to the temperature difference.

Clearly, the two displacements $r_{12}$ and $r_{23}$ in Eq.~(\ref{r12r23}) correspond 
to the sphere positions in their model.
It is interesting to note, for example, that our result of the frequency matrix in 
Eq.~(\ref{frequencymatrix}) reduces to Eq.~(42) in Ref.~\cite{Gnesotto18} when $T_2=0$.
When $T_2 \neq 0$, however, the presence of the middle sphere changes the structure 
of the frequency matrix as we have shown in Eq.~(\ref{genfrequencymatrix}) or 
Eq.~(\ref{frequencymatrix}).
Moreover, the important outcome of this work is the relation between the average velocity
and the eigenvalue of the frequency matrix for a three-sphere micromachine [see 
Eq.~(\ref{Vgamma1}) or Eq.~(\ref{Vgamma2})]. 
Notice that a two-sphere micromachine in a viscous fluid cannot have a directed motion 
even if the temperatures are different~\cite{Hosaka17}.

As mentioned before, we have neglected long-ranged hydrodynamic interactions acting 
between different spheres.
In our previous paper~\cite{Hosaka17}, we explicitly took into account hydrodynamic interactions
when the friction coefficients are all identical.
If hydrodynamic interactions are taken into account in the present analysis, the variances and 
the covariance in Eqs.~(\ref{sigma12})--(\ref{sigma13}) are modified in non-equilibrium situations.
Such hydrodynamic corrections should be proportional to $a/\ell$ within the lowest-order 
expansion.
Moreover, such corrections should vanish in thermal equilibrium, i.e., $T_1=T_2=T_3$ because 
hydrodynamic interactions should not affect equilibrium statistical properties.

In the presence of hydrodynamic interactions, a thermally driven elastic micromachine can 
undergo a directional motion~\cite{Hosaka17}.
The presence of the middle sphere is essential for a directional motion because the 
hydrodynamic interactions among the three spheres are responsible for it.
Such a locomotion should be distinguished from the traditional thermophoresis (Ludwig-Soret 
effect) in which a temperature gradient in an external fluid induces a directed motion of suspended 
particles~\cite{Jiang10}.
In our model, the locomotion of a micromachine is purely induced by non-equilibrium fluctuations 
of internal degrees of freedom.
Here, the three spheres are in thermal equilibrium with independent heat baths having different 
temperatures, which is different from a situation where a temperature gradient is externally 
imposed in a surrounding fluid~\cite{Jiang10}. 
For example, we showed before both analytically and numerically that a two-sphere elastic 
micromachine cannot move even if the temperatures are different~\cite{Hosaka17}.
This clearly indicates that the locomotion of a thermally driven micromachine cannot be explained 
within the standard thermophoresis, although it is closely related to Purcell's scallop theorem for 
microswimmers under the force-free condition.

In the present work, we have shown that asymmetries in the friction coefficients lead 
to a finite probability flux loop [see Fig.~\ref{flux}(d)] or flux rotor [see Fig.~\ref{rotation}(d)], 
anticipating a locomotion of an asymmetric micromachine in a viscous fluid.
This prediction will be investigated in the future by performing numerical simulations of the coupled 
stochastic equations, as given by Eqs.~(\ref{x1dot})--(\ref{x3dot}), in the presence of hydrodynamic 
interactions.
Notice that analytical treatment of a case with different sphere sizes is more difficult because one 
needs to take into account higher-order contributions in $r_{12}$ and $r_{23}$ as well as in the sphere 
size $a$.

So far, various concepts have been proposed to quantitatively discuss whether a steady state is 
in thermal equilibrium or not~\cite{Rupprecht16}.
One of the most promising methods is to search for the violation of the fluctuation-dissipation 
relation that is guaranteed in thermal equilibrium situations~\cite{Guo14,Turlier16}.
However, it is not always easy to perform two separate measures of the correlation function and 
of the response function for the same system.
Moreover, the measurement of the response function can often be intrinsically invasive and 
is not suitable for biological systems.  
Moreover, the observation of non-Gaussian distribution fluctuations is not a proof for the 
non-equilibrium steady sate~\cite{Bustamante05,Battle16}. 
Since the emergence of probability flux loops is a direct verification of a non-equilibrium steady 
state, it can be a powerful method to study systems that are driven out of 
equilibrium.
We consider that a thermally driven three-sphere micromachine is an excellent example to show 
the usefulness of
such an analysis.

\acknowledgements

We thank T.\ Kato and  M.\ Doi for useful discussions.
We also thank the anonymous referee for the useful suggestions.
Y.H.\ acknowledges support by a Grant-in-Aid for JSPS Fellows (Grant No.\ 19J20271) from the Japan Society 
for the Promotion of Science (JSPS). 
K.Y.\ acknowledges support by a Grant-in-Aid for JSPS Fellows (Grant No.\ 18J21231) from the JSPS.
S.K.\ acknowledges support by a Grant-in-Aid for Scientific Research (C) (Grant No.\ 18K03567 and
Grant No.\ 19K03765) from the JSPS.

\appendix*
\section{Matrix representation of stochastic dynamical systems}
\label{sec:formulation}

Following closely the argument in Refs.~\cite{Weiss03,Weiss07}, we shall briefly review the general 
formulation of stochastic dynamical systems using a matrix representation. 
Let us start from the following linear stochastic Langevin model
\begin{align}
\dot{{\bm r}} = {\mathbf A} {\bm r}  + {\bm \xi},
\label{genlangevin}
\end{align}
where ${\bm r}$ is a $N$-dimensional vector characterizing the state of the system, 
${\mathbf A}$ is a $N\times N$-dimensional matrix describing the linear deterministic dynamics, 
and ${\bm \xi}$ is a $N$-dimensional vector representing the noise forcing.  
The time-dependent covariance matrix is introduced by 
${\mathbf C} = \langle {\bm r}(t) {\bm r}^{\mathsf T}(t') \rangle$.
Then the diffusion matrix ${\mathbf D}$ is defined by the relation
\begin{align}
\langle {\bm \xi}(t) {\bm \xi}^{\mathsf T}(t') \rangle = 2{\mathbf D} \delta(t-t').
\label{diffusion}
\end{align}
From Eq.~(\ref{genlangevin}), one can show that the time evolution of the covariance matrix is given by 
\begin{align}
\dot{{\mathbf C}} = {\mathbf A}{\mathbf C} + {\mathbf C}{\mathbf A}^{\mathsf T} + 2 {\mathbf D}.
\label{timeevolution}
\end{align}

When the system is in a steady state, i.e., $\dot{{\mathbf C}}=0$, the covariance matrix must 
obey the following Lyapunov equation:
\begin{align}
{\mathbf A}{\mathbf C} + {\mathbf C}{\mathbf A}^{\mathsf T} + 2 {\mathbf D}=0.
\label{lyapunov}
\end{align}
Hereafter, we use the same notation ${\mathbf C}$ for the covariance matrix that satisfies the Lyapunov equation.
It should be emphasized that the Lyapunov equation holds both in equilibrium and out of equilibrium situations.
Hence it can be regarded as a generalized fluctuation-dissipation relation connecting 
the fluctuations ($\mathbf C$) and the deterministic dissipation ($\mathbf A$).

For linear systems with Gaussian noise, the steady state probability distribution function 
$p$ is a Gaussian function as in Eq.~(\ref{gaussian}).
In this case, the probability flux defined by Eq.~(\ref{probflux}) becomes 
\begin{align}
{\bm j} = {\mathbf A} {\bm r} p - {\mathbf D} \nabla p
=({\mathbf A}+ {\mathbf D} {\mathbf C}^{-1}) {\bm r} p,
\end{align}
Hence the frequency matrix ${\bm \Omega}$ introduced through the relation 
${\bm j} ={\bm \Omega} {\bm r}p$ [see Eq.~(\ref{jomega})] is given by 
\begin{align}
{\bm \Omega} ={\mathbf A}+ {\mathbf D} {\mathbf C}^{-1}.
\label{Omega}
\end{align}

When ${\bm \Omega}=0$ and hence detailed balance holds, we have 
${\mathbf A}+ {\mathbf D} {\mathbf C}^{-1}=0$.
After the elimination of ${\mathbf C}$ in the Lyapunov equation, we then have 
\begin{align}
{\mathbf A}{\mathbf D} - {\mathbf D}{\mathbf A}^{\mathsf T} =0.
\label{detailed}
\end{align}
This commutation relation holds if and only if the system is in thermal equilibrium and 
satisfies detailed balance.
Moreover, whether or not detailed balance is satisfied is coordinate invariant~\cite{Weiss03}.

When ${\bm \Omega} \neq 0$ and hence detailed balance is broken, the system is in 
non-equilibrium steady state situations.
In this case, one can show that the following two relations hold: 
\begin{align}
{\mathbf A}{\mathbf C} + {\mathbf D} ={\bm \Omega} {\mathbf C},~~~~~
{\mathbf C}{\mathbf A}^{\mathsf T} + {\mathbf D} = -{\bm \Omega} {\mathbf C}.
\label{twodetailed}
\end{align}
Obviously, the sum of these two relations gives again the Lyapunov equation in 
Eq.~(\ref{lyapunov}).
One can also show that the flow field, as expressed by ${\bm \Omega} {\bm r}$, is 
perpendicular to the gradient of the probability distribution, i.e., 
$({\bm \Omega} {\bm r}) \cdot \nabla p=0$~\cite{Gnesotto18}.

Last but not least, the matrices ${\mathbf C}$ in Eq.~(\ref{covarianceC}), 
${\mathbf A}$ in Eq.~(\ref{Amatrix}), ${\mathbf D}$ in Eq.~(\ref{Dmatrix}), 
and ${\bm \Omega}$ in Eq.~(\ref{genfrequencymatrix}) obtained for a three-sphere
micromachine satisfy all the relations in this Appendix.


\end{document}